**Neonatal sepsis is diminished by cervical vagus nerve stimulation and tracked non-invasively by ECG: a pilot report and dataset in the piglet model**

Aude Castel[1*], P. Burns[1*], C. Wakefield[4], K. J. Jean[3], Y. Frank[4], M. Cao[3], A. Desrochers[1], G. Fecteau[1], C. Faure[5], C. L. Herry[2], M.G. Frasch[3,4,6,7]

* equal contribution

*[1]Dept. of Clinical Sciences, Faculty of Veterinary Medicine, Université de Montréal, QC, Canada*
*[2]Ottawa Hospital Research Institute, University of Ottawa, ON, Canada*
*[3]Dept. of Obstetrics and Gynaecology and Dept. of Neurosciences, CHU Ste-Justine Research Centre, Faculty of Medicine, Université de Montréal, Montréal, QC, Canada*
*[4]Dept. of Obstetrics and Gynaecology, University of Washington, Seattle, WA, USA*
*[5]Dept. of Pediatrics, CHU Ste-Justine Research Centre, Faculty of Medicine, Université de Montréal, Montréal, QC, Canada*
*[6]Animal Reproduction Research Centre (CRRA), Faculty of Veterinary Medicine, Université de Montréal, Montréal, QC, Canada;*
*[7]Center on Human Development and Disability, University of Washington, Seattle, WA, USA*

**Address of correspondence:**
Martin G. Frasch
Department of Obstetrics and Gynecology
University of Washington
1959 NE Pacific St
Box 356460
Seattle, WA 98195
Phone: +1-206-543-5892
Fax: +1-206-543-3915
Email: mfrasch@uw.edu

## Abstract

**Background**: Vagus nerve stimulation (VNS) reduced inflammation induced by lipopolysaccharide (LPS) in an adult rat sepsis model. A multi-dimensional heart rate variability (HRV) index reliably tracked the inflammatory profile in near-term sheep fetuses. The effects of VNS on neonates are not known. First, in a neonatal piglet model of sepsis, we present an approach to evaluate the effect of VNS on systemic inflammatory response induced by a high dose of LPS to mimic late-onset neonatal sepsis. Second, we present an analytical pipeline to validate our fetal-sheep-derived HRV inflammatory index in neonatal piglets to test its performance in different species, older developmental stages, and more robust septic responses.
**Methods**: Three neonatal piglets of 7–14 days of age with 2.4–4 kg in body weight were used in this proof-of-principle study. Following anesthesia, electrodes were attached bilaterally to the cervical portion of the vagus nerve to allow for stimulation of the left vagus (VNS) and bilateral vagus electroneurogram (VENG). Electrocardiogram (ECG), blood pressure (BP), and VENG were recorded for the duration of the experiment. After baseline recording, the piglets were administered LPS at 2mg/kg IV bolus. In the VNS-treated piglet, the vagus nerve was stimulated for 10 minutes prior to and 10 min after the injection of LPS. In both groups, each 15 min post LPS, an arterial blood sample was drawn for blood gas, lactate, and glucose as well as the inflammatory cytokines measured by a quantitative ELISA multiplex panel. At the end of the experiment, the piglets were euthanized. BP and ECG-derived HRV were calculated and VENG was analyzed.
**Results**: The piglets developed a potent inflammatory response to the LPS injection with TNF-α, IL-1β, IL-6 and IL-8 peaking between 45 and 90 min post-injection. VNS diminished the LPS-induced systemic inflammatory response, with a decrease in measured cytokines levels ranging from two to ten-fold. We present a low-cost, easy-to-implement design of a VNS/VENG probe and a framework to analyze VENG in response to LPS. Furthermore, the HRV index accurately tracked cytokine temporal profiles' which was reflected in the power-spectral and complex properties of VENG when applying the machine learning model derived from HRV.
**Discussion**: We present a novel method to model, manipulate and track neonatal sepsis using VNS/VENG. Our supportive findings suggest that 1) the HRV index of the systemic inflammatory response applies across species pre- and postnatally, 2) the HRV index performs well at different degrees of sepsis (i.e., nanogram and milligram doses of LPS), and 3) the present VNS paradigm effectively suppresses LPS-induced inflammation, even with high doses of LPS;  an effect that is reflected by changes in the shared mathematical properties of both VENG and HRV. These findings suggest that the HRV inflammatory index reflects underlying changes in the VENG activity. The presented method lays a foundation for larger studies to investigate the mechanisms and therapeutic potential of early postnatal VNS intervention to counteract sepsis progression. Moreover, the presented experimental and analytical frameworks highlight the potential for HRV monitoring to serve as an early biomarker for tracking the systemic inflammatory response.

## Introduction

Sepsis is a life-threatening yet treatable – when detected in time – condition of global significance.[1] [1] There are an estimated 30 million episodes of sepsis annually, with neonates having the highest incidence of any age group (~3 million babies annually). The burden of neonatal sepsis is profound, carrying an associated mortality rate of 11–19%.[2] Although the majority of children do recover, many suffer lifelong sequelae following the inflammatory response. Given the particularly vulnerable nature of neonates, early detection and prevention of sepsis will have a profound impact on reducing overall morbidity and mortality.

Lipopolysaccharide (LPS) is commonly used to elicit and study septic inflammatory responses as it induces endotoxemia mimicking an infection of the blood by Gram-negative bacteria. The pig is considered an excellent model for septicemia, necrotizing enterocolitis and neonatal brain injury due to its anatomical and physiological similarities with humans. As a means of modulating the septic inflammatory response, vagus nerve stimulation (VNS) has been shown to exert anti-inflammatory and anticoagulant responses induced by LPS in an adult rat sepsis model.[3]

As a primary objective, we aimed at evaluating the effect of stimulating the cholinergic anti-inflammatory pathway (via stimulation of the vagus nerve) on the systemic (plasma cytokines) inflammatory response induced by an IV injection of a high dose of LPS.

We have shown that multi-dimensional heart rate variability (HRV) analysis can reliably track the inflammatory profile in near-term sheep fetuses (Fig. 1). As a secondary objective, we aimed at validating the derived HRV inflammatory index in a different setting of sepsis and at comparing its performance and temporal profile with the direct analysis of the underlying vagus nerve electrical activity (vagus nerve electroneurogram, VENG).

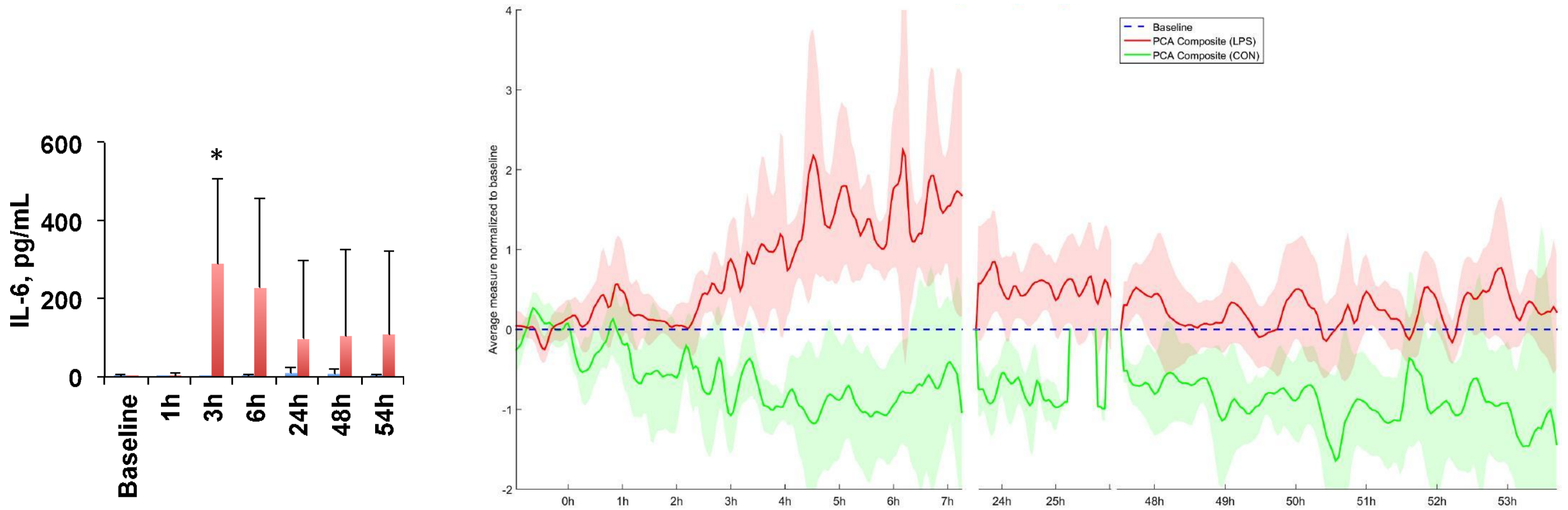

**Figure 1. Heart rate variability (HRV) provides a signature of the fetal systemic inflammatory response in a fetal sheep model of lipopolysaccharide-induced sepsis. LEFT:** Temporal profile of LPS-induced fetal sheep inflammatory response measured by plasma levels of IL-6 with a peak at 3 hours post fetal i.v. LPS injection - also reflected in the fHRV inflammatory signatures. **RIGHT:** Principal component analysis (PCA) approach to identify fHRV signatures which track IL-6 responses. From [4] with permission.

## Materials and Methods

### Ethics Statement

This pilot study was carried out in strict accordance with the recommendations in the Guide for the Care and Use of Laboratory Animals of the National Institutes of Health. The respective *in vivo* protocol was approved by the Committee on the Ethics of Animal Experiments of the Université de Montréal (Permit Number: 13-Rech-1695).

### Anesthesia

Three neonatal piglets of 7–14 days of age with 2.4–4 kg body weight were used in this pilot study (see Table 1). Each piglet was pre-medicated with butorphanol (analgesic) (0.1mg/kg) and diazepam or midazolam (sedative) (0.2mg/kg) intramuscularly, approximately 15 minutes prior to anesthesia. Anesthesia was induced using isoflurane in oxygen via a mask (expired isoflurane 1 to 2.5%; ETISO%). The piglet was then intubated and mechanically ventilated (expired $CO_2$ = 35 to 45 mm Hg; $ETCO_2$). Monitoring included capnography, direct arterial blood pressure (ABP), central venous pressure, pulse oximetry, electrocardiography (ECG) and temperature. Body temperature was maintained using warm water blankets. Observations were taken every five minutes.

### Surgery

Once the piglet was anesthetized, the catheter insertion sites (neck and groin) were surgically prepared. A 20 to 22 G catheter was inserted into the femoral artery in the groin area via a cut-down to measure the direct ABP. An introducer catheter was inserted into the right carotid via a cut-down. A second catheter was then inserted through the introducer catheter and fed into the left ventricle. The correct placement of the catheter was verified using the arterial pressure trace. Lidocaine 2% was instilled (splash block) into the surgical wounds prior to these first two procedures. Another catheter (8 to 10 Fr G) was inserted into the left jugular via a cut-down to measure the central venous pressure and to administer intravenous fluids (3 to 5 ml/kg/hr). Via an incision on each side of the neck, electrodes were attached to the left and right vagus nerves to allow for stimulation of the left vagus nerve and bilateral recording (VENG). Left VNS has been shown to result in less to no cardiovascular side effects, *i.e.*, no bradycardia.[5,6] In the control animal (LPS only), the electrodes were also placed on both vagus nerves in the same way to record VENG (but no VNS). Local anesthesia was not used in this surgical incision because of the risk of desensitizing the vagus nerve. A suprapubic urinary catheter was inserted into the urinary bladder to measure the urine output and to minimize the discomfort associated with a distended bladder during the experiment (~3 hours postoperatively). A small incision (3 to 4 cm) was made to facilitate the insertion of this urinary catheter. A purse-string suture was placed to secure the urinary catheter in place.

### Data acquisition

ECG, heart rate (HR), and ABP were monitored continuously (1902 amplifier and micro3 1401 ADC by CED, Cambridge, U.K., and NL108A, NeuroLog, Digitimer, Hertfordshire, U.K) and sampled at 1000 and 256 Hz, respectively. VNS was applied via NeuroLog's NL512/NL800A using a pulse sequence pre-programmed in Spike 2. The VNS settings were as follows: DC rectangular 5 V, 100 µA, 2 ms, 1 Hz according to [7]. VENG was recorded at 20,000 Hz. See Fig. 2 for VNS/VENG electrode design.[8]

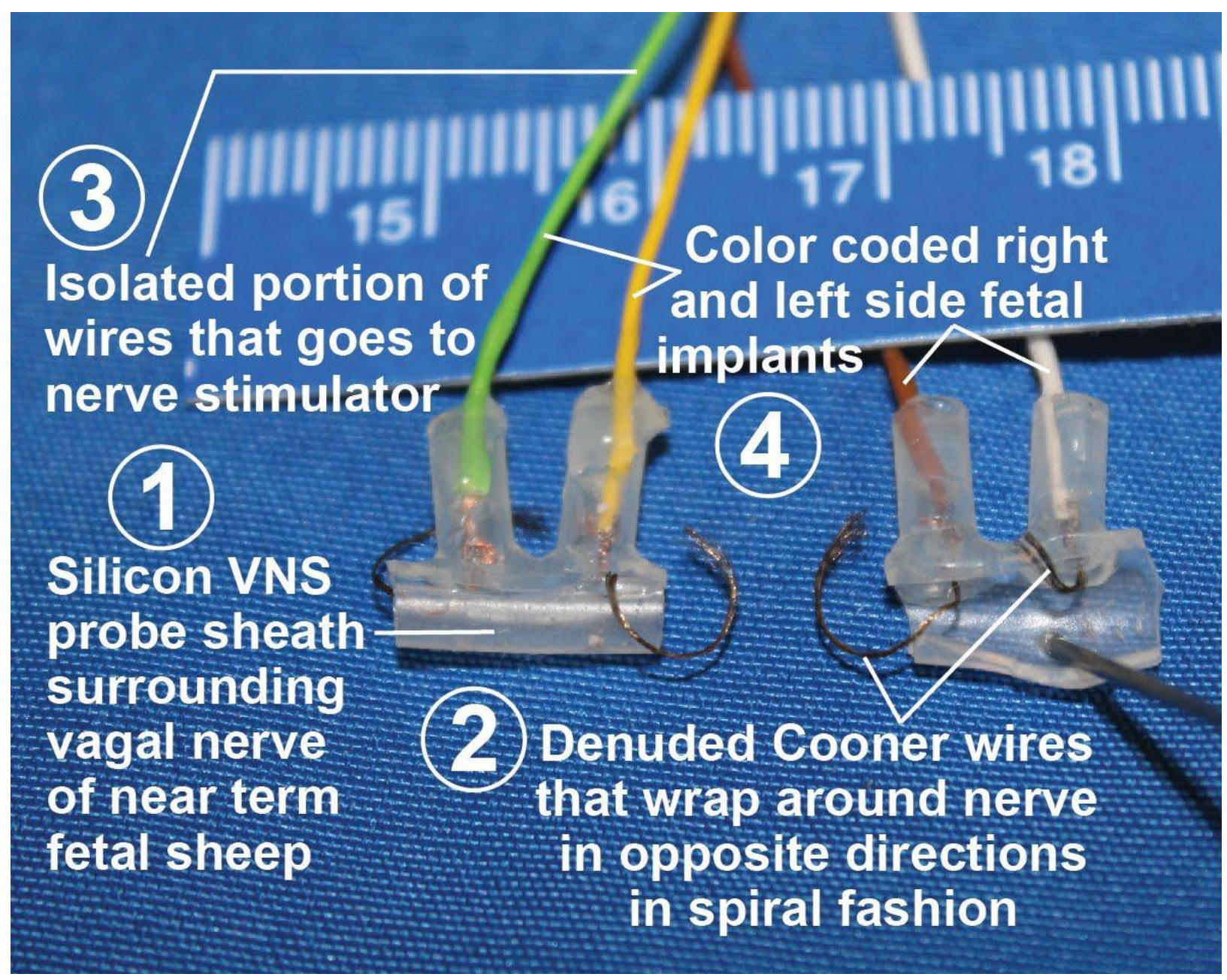

**Figure 2. Design of the fetal sheep cervical bilateral VNS/VENG probe.**[8] The scale is metric. This design was also used successfully in the present approach in the neonatal piglet.

### Experimental protocol

*Baseline*

After a period of stabilization (15 minutes), 60 min of "baseline" recording was allowed during which ABP, ECG and VENG were recorded. In the VNS+LPS piglet, only the left VENG was recorded. In the LPS piglet, bilateral VENG was recorded. Subsequently, a baseline blood sample was taken for blood gases (Radiometer, 0.8 ml), complete blood count (hematology) (1cc) and cytokines ELISA (3cc, spun down at 4° C, 4000 rpm for 4 minutes and frozen at -80 °C for plasma) (total of 5 mL blood).

*Induction of endotoxemia with LPS*

Next, the piglets were administered LPS 2mg/kg IV bolus (Sigma L2880).[9] In the treatment group (VNS), the vagus nerve was stimulated for 10 minutes prior to and 10 min after the injection of LPS. The control group was observed over the course of 3 hours as it developed sepsis. In both groups, every 15 min post LPS administration, a 0.8 ml arterial blood sample was drawn for blood gas, lactate, and glucose (Radiometer ABL800 Flex). At baseline, 15, 45, 90 and 135 minutes, these measurements were done as part of the larger blood sample (3–4 ml) together with inflammatory cytokines and for hematology. At the end of the experiment, the piglets were euthanized with an overdose of pentobarbital and tissues were collected. Prior to the injection of pentobarbital, the level of anesthesia was deepened.

*Cytokines assay*

Plasma cytokines were measured using the commercial service provided by Eve Technologies (Calgary, AB, Canada). Porcine Cytokine 13-plex Discovery Assay kits were used (Cat.# PCYTMAG-23K-13PX) to examine 13 cytokines including GM-CSF, IFN-γ, IL-1α, IL-1β, IL-1ra, IL-2, IL-4, IL-6, IL-8, IL-10, IL-12, IL-18, and TNF-α, Here we report the results of IL-1β, IL-6, IL-8, IL-10 and TNFα, the cytokines of most interest mediated by the cholinergic anti-inflammatory pathway[10]; their assay sensitivities are 42 pg/ml, 9 pg/ml, 5 pg/ml, 9 pg/ml and 6 pg/ml, respectively. The intra-assay and inter-assay variations for all the cytokines were <10% and < 20%, respectively.

*Hematology*

The extent of inflammation was further evaluated by using a complete blood count (CBC) to compare the hematological changes associated with sepsis in both groups.

*Data analysis*

Evaluation of piglets' physiology was done based on the literature.[11,12] Mean (mBP), diastolic (dBP) and systolic (sBP) ABP, as well as HR, were calculated for each animal, at each time point, as an average of the artifact-free 10 preceding minutes (60 preceding minutes for the baseline) using Spike 2 (Version 7.13, CED, Cambridge, U.K.). We reported the CIMVA approach and derivation of the HRV composite measure elsewhere.[4,13] The VENG analysis was conducted with the open-source EEGLAB package v2019_1 within the Matlab environment (Matlab 2013b for Linux, MathWorks, Natick, MA). The power spectral analysis was done in Python. The dataset and all code scripts are available on FigShare.[14]

## Results

The animals' characteristics are summarized in Table 1. LPS injection had a rapid and profound effect on piglets' acid-base status characterized by rising lactic acidosis which was antagonized by VNS treatment (Fig. 3).

***Table 1. Animal age, gender and body weight, and treatment***

| Animal ID | Age (days) | Gender | Bodyweight (kg) | Treatment |
|---|---|---|---|---|
| Piglet 1* | 7 | Male | 2.5 | Control |
| Piglet 2 | 7 | Male | 2.4 | LPS + VNS |
| Piglet 3 | 14 | Male | 4 | LPS |

* lost during surgical instrumentation due to a complication with intubation

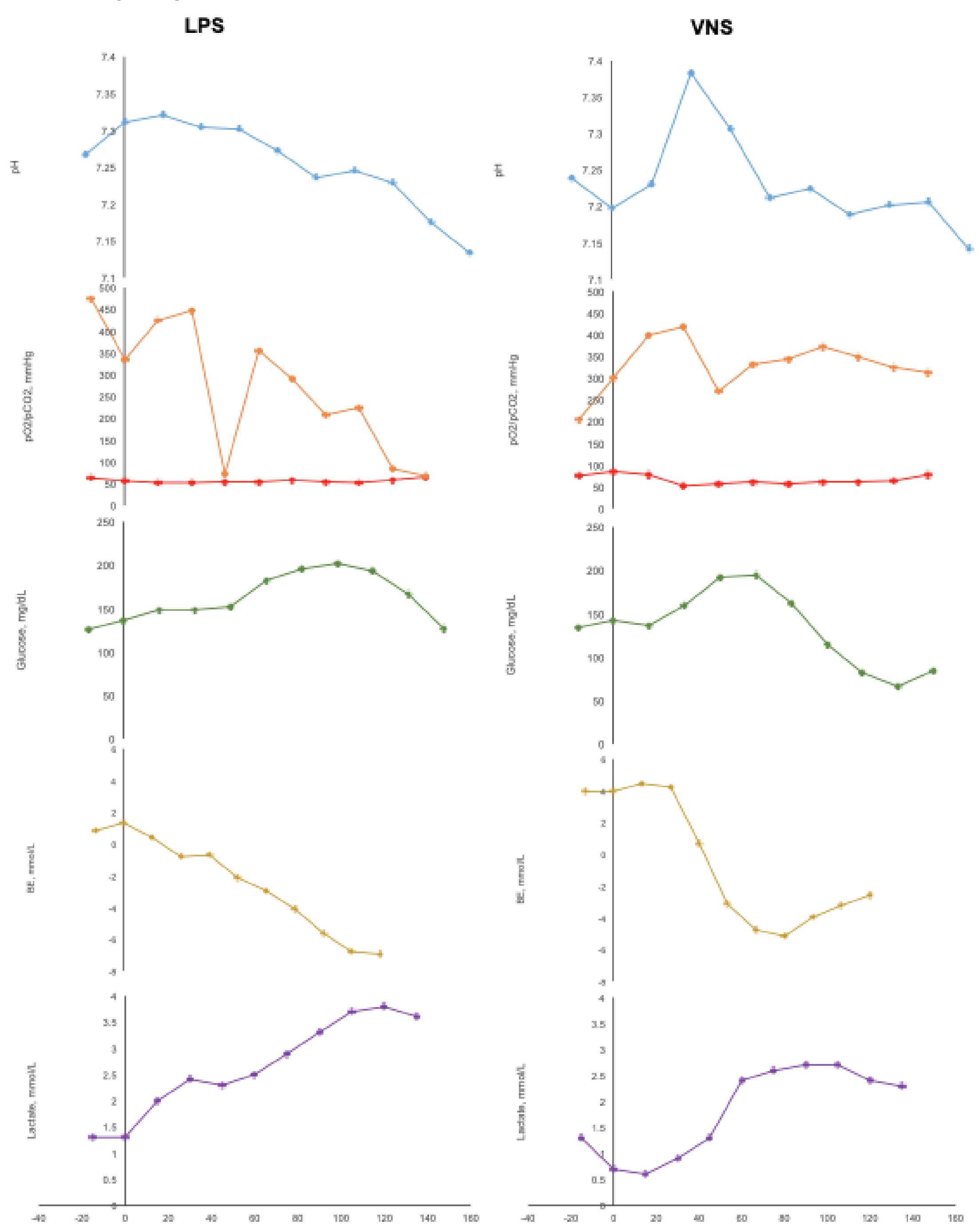

**Fig. 3. Blood gas, glucose and lactate measurements during the experiment.**
Piglet 3 (LPS) developed profound and progressive leukopenia followed by the appearance of toxic neutrophils toward the end of the experiment at 135 minutes. Piglet 2 (LPS+VNS) developed toxic neutrophils after 135 minutes.

### Cardiovascular effects

In piglets, HR of less than 120 is considered bradycardia, whereas between 121 and 160 is normal, and more than 161 represents tachycardia. The VNS-treated animal showed a more stable, albeit tachycardic HR of around 220 bpm, while LPS alone resulted in a rapid rise in HR to 240 bpm, following the temporal profile of the cytokine increases toward 100 min post LPS (Fig. 4 and Fig. 5). The VNS-treated piglet started out with normal sBP but lower than physiological levels of mBP and dBP, immediately following LPS injection and maintained sub-physiological mBP and dBP even at the time of TNF-α peak around 45 min post LPS injection. This difference disappeared toward 80 min post LPS and in the further course of the experiment.

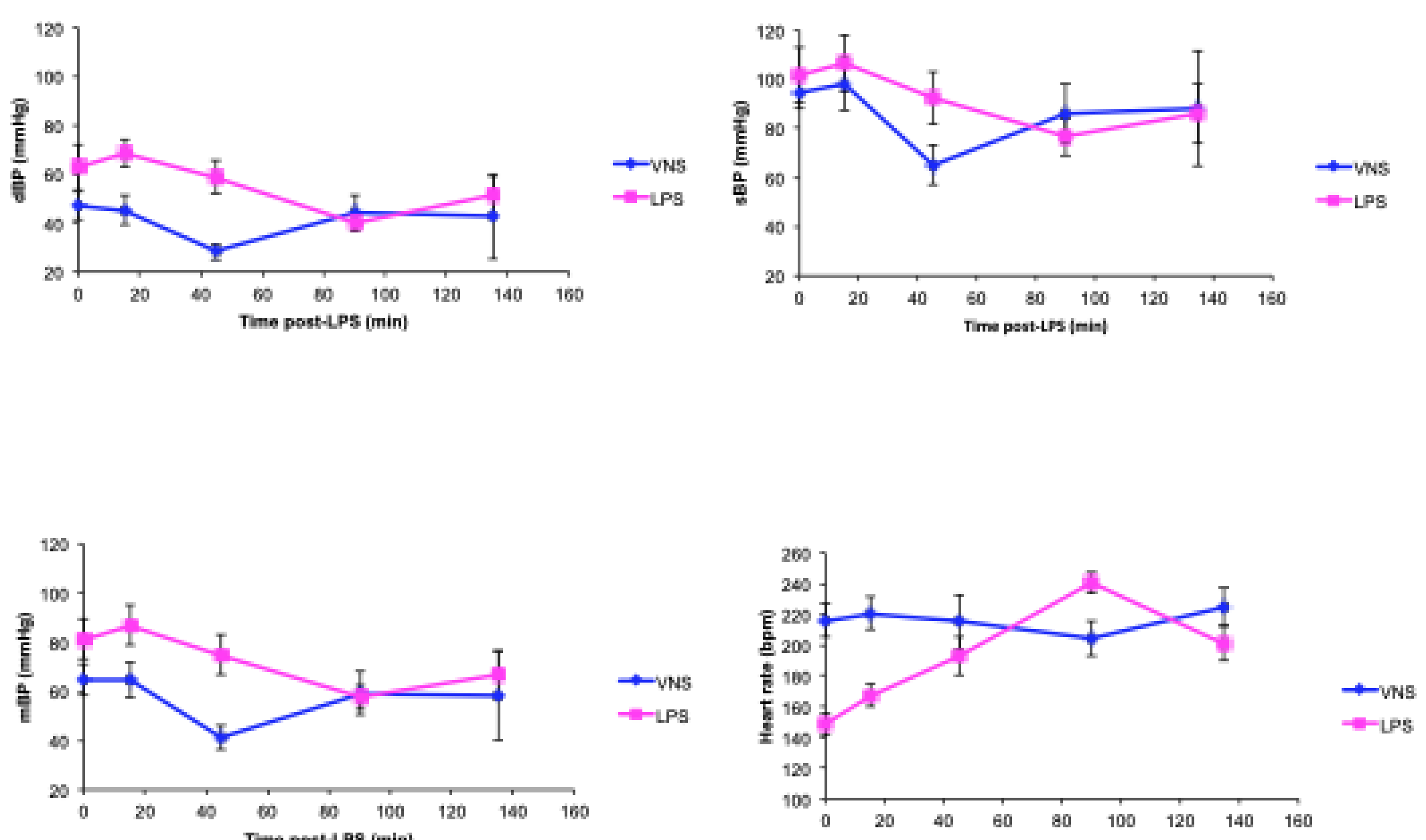

**Fig. 4. Cardiovascular responses: diastolic, systolic and mean blood pressure as well as heart rate.**

### Cytokine responses

At 45 min post LPS injection, TNF-α levels peaked and were eight-fold lower in the VNS-treated animal (Fig. 5). IL-10 and IL-6 followed the same trend reflecting the physiological compensatory rise of the anti-inflammatory IL-10. Surprisingly, VNS also limited the rise of IL-10 at this time point, delaying its increase by ~90 min. IL-6 level responded with a characteristic slight delay in rising at ~90 min and VNS reduced the magnitude of this response by two-fold.[15,16] There was no effect of VNS on the LPS-induced increases of IL-1β and IL-8. It is possible that the suppression of the IL-10 rise by VNS secondarily abolished its effect on IL-1β. Not shown in Fig. 5, we also observed an increase of IL-4, IL-12, and IL-18 following LPS administration. VNS treatment accelerated the rise of IL-4 and IL-12 by 45 minutes to a similar peak magnitude of 200 pg/mL and 1200 pg/mL at 135 min, respectively. The IL-18 level showed a biphasic temporal profile rising with a 45 min delay to a similar level of 425 pg/mL in the VNS-treated animal.

### HRV behavior

HRV inflammatory index tracked the inflammatory response and its change secondary to VNS closely over time, particularly following the trend of the temporal profile of TNF-α at its primary peak around 45 min and showing a smaller, secondary peak at ~90 minutes which corresponded to the delayed peaking levels of IL-6, again reduced by VNS (Fig. 5). Due to some missing ECG data in this segment of the experiment, there were corresponding missing time points in the HRV index.

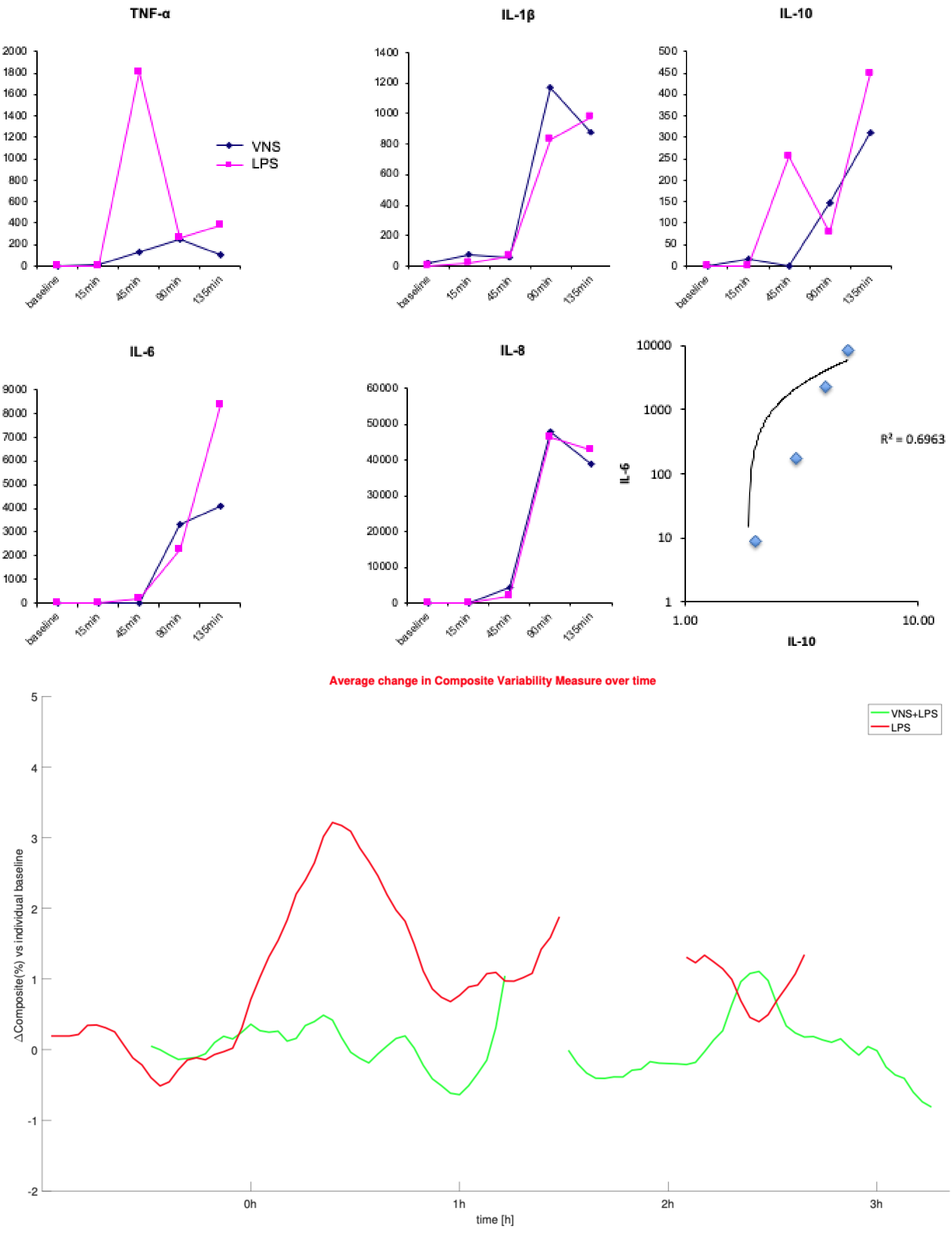

**Fig. 5. Cytokines and HRV. Validation of fHRV signature of inflammation (derived from the near-term fetal sheep model of low-dose LPS-induced fetal inflammatory response) in a neonatal piglet model of sepsis at a higher LPS dose**. **TOP**: Inflammatory response to LPS injection in 2 piglets receiving each 2 mg/kg intravenous LPS after baseline with or without vagus nerve stimulation (VNS). VNS diminished LPS-induced systemic cytokine production (pg/mL). **BOTTOM**: The HRV composite measure derived from the fetal sheep model of low-dose intravenous LPS exposure for tracking the inflammatory response is applied to this piglet model following high-dose intravenous LPS exposure. Note that the HRV composite measure tracked accurately the cytokines' change over time as seen by comparing the peaks and troughs with the timing on the X-axes in each diagram).

### VENG patterns in response to LPS and VNS

VENG power spectra over the course of the experiment are presented in Figure 6. Since most changes are visible above 2 kHz, we visualized the data in several ways to demonstrate this behavior more clearly. To achieve this, we determined spectral power in low (1–1000 Hz), medium (1000 – 2000 Hz), and high frequency (2000–10000 Hz) bands of VENG activity, in addition to the whole-band power spectral analyses.

### *Piglet receiving LPS only*

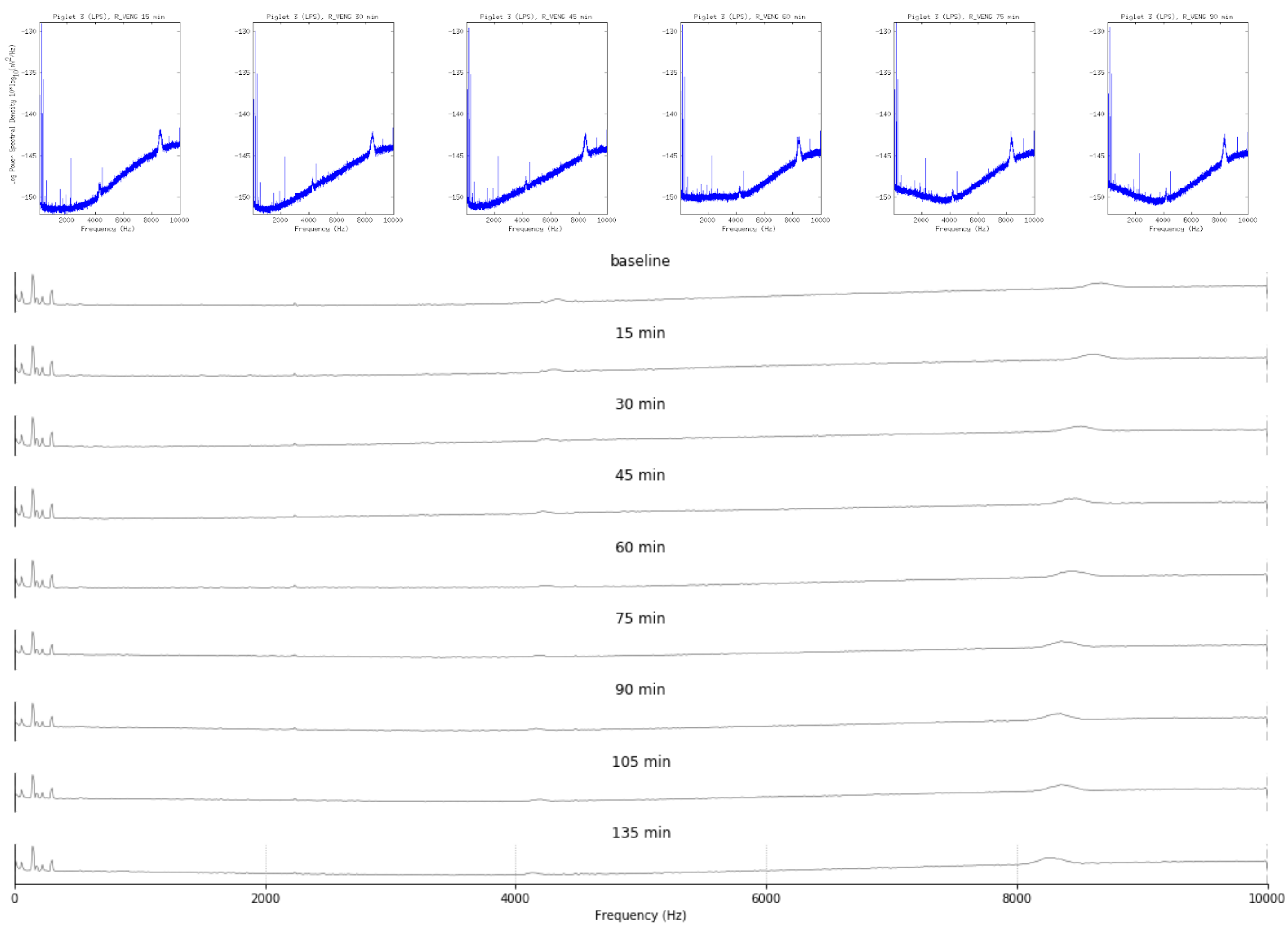

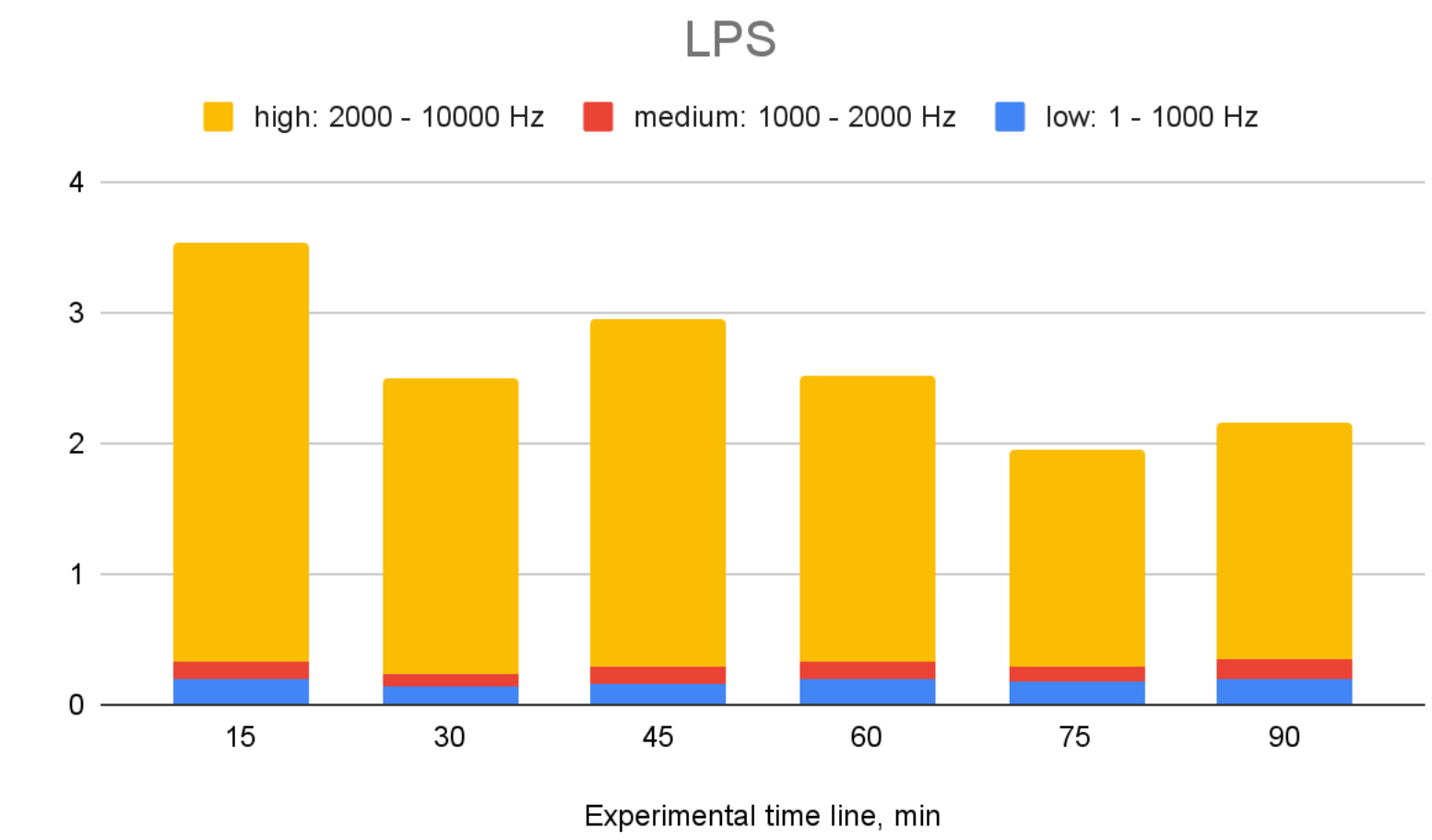

***Piglet receiving VNS treatment in addition to LPS (LPS+VNS)***

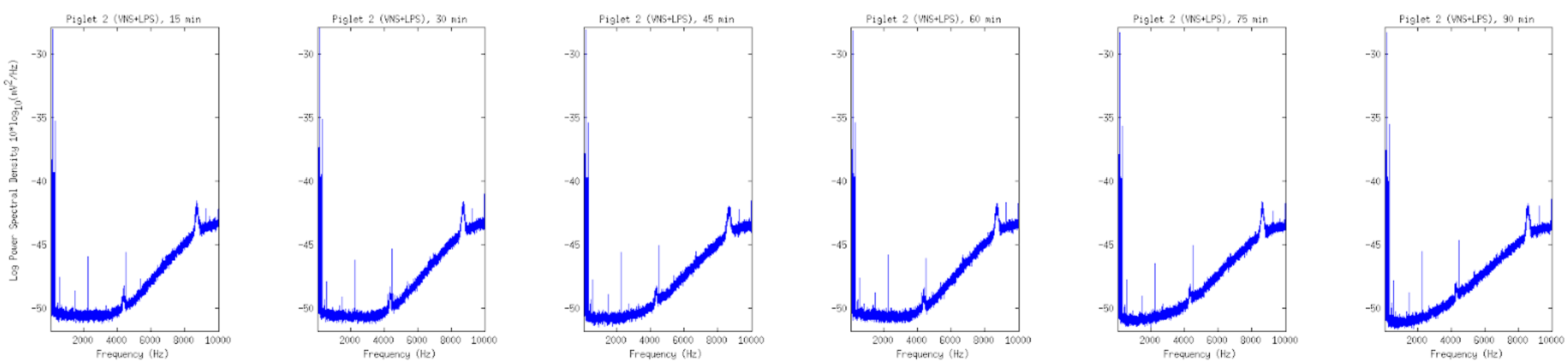

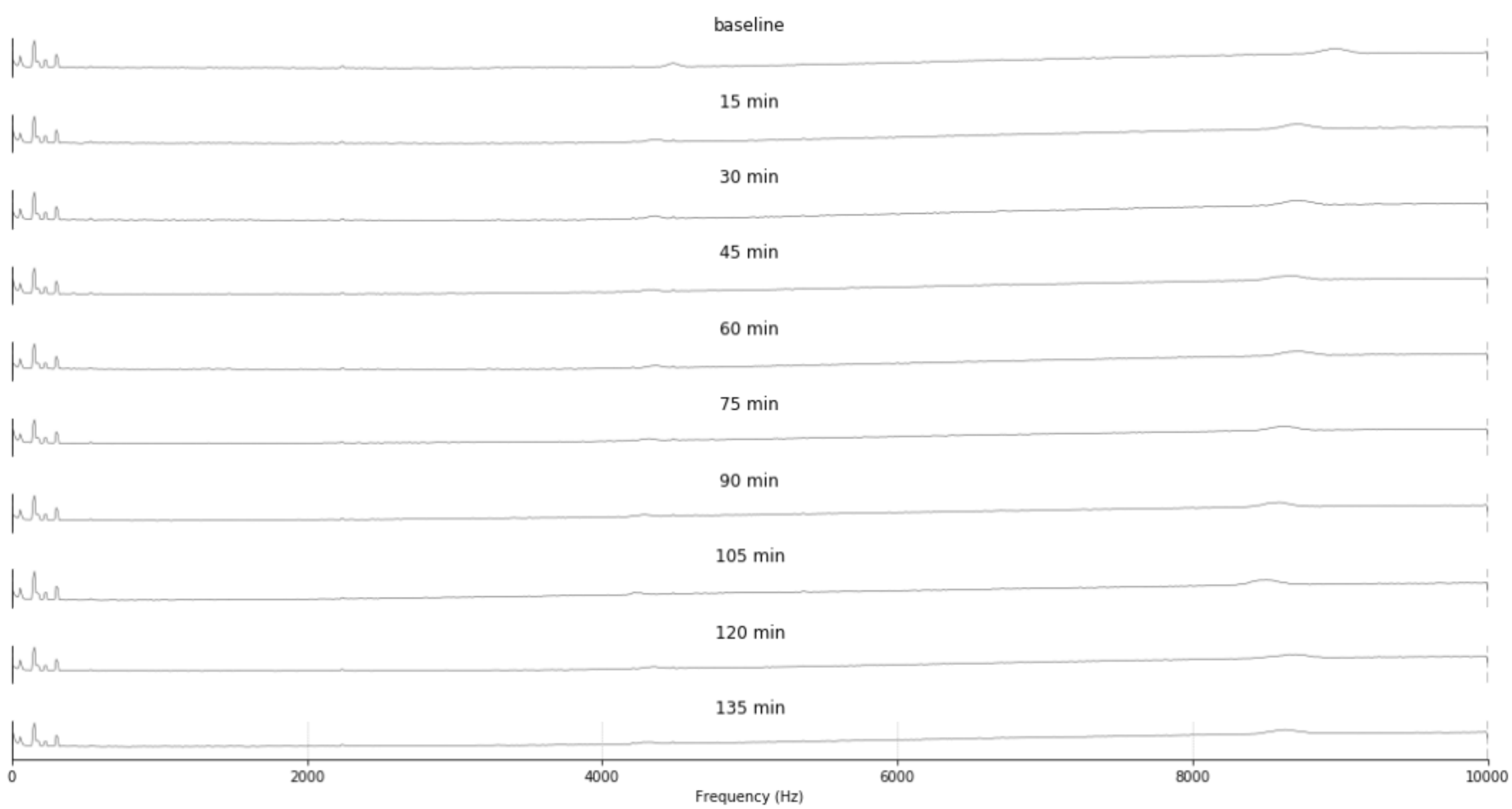

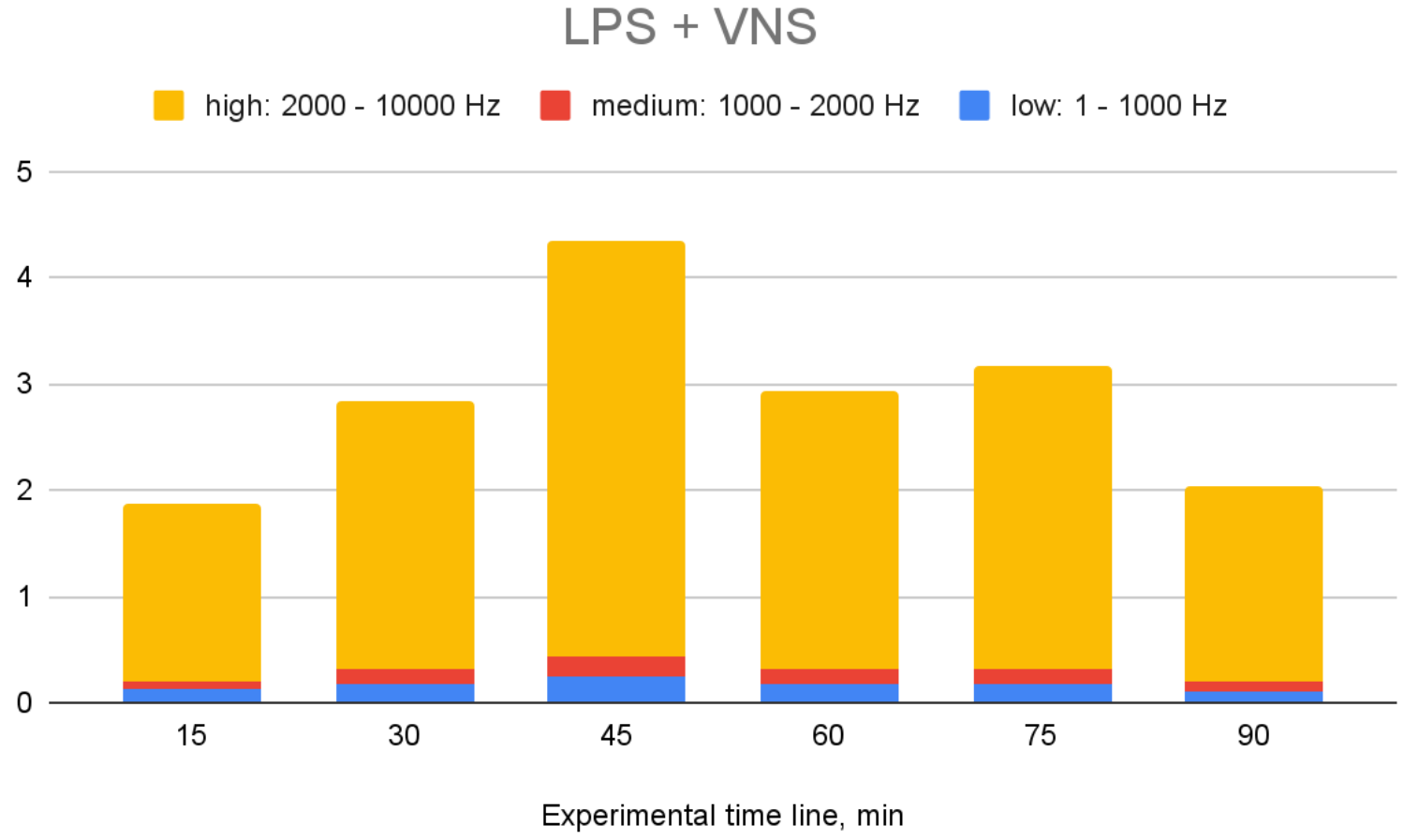

**Fig. 6. Temporal frequency dynamics of VENG displayed as power spectral density and summated within sub-frequency bands.** Note that most changes occur above 2,000 Hz.
Data is shown from one neonatal piglet exposed to LPS without VNS (TOP) or another piglet with VNS treatment (BOTTOM). Y-axis ($mV^2$/Hz) is optimized for each series to highlight the frequency peaks.

We observed several immediate differences between the VENG of the LPS+VNS piglet compared to the LPS-only piglet. First, the VENG of the LPS+VNS piglet showed 3-fold higher average levels of VENG spectral power density over the course of the experiment compared to the piglet exposed to LPS alone.

Second, there was a change in the frequency peaks distribution across the spectrum. VENG showed several common as well as VNS-treatment-specific spectral frequency patterns. The common patterns included the recurring and time-specific peaks in the ranges of 4.1–4.2 and 8.2–8.3 kHz. VNS moved these peaks to the right of the power spectrum by about 200 Hz.

Third, the slope of the rise of the power spectral density changed with the time course of the experiment following LPS administration. The rise of the slope began earlier in time and started at a lower frequency in the LPS-only piglet's VENG compared to VNS–treated piglet (2500 Hz at ~30–45 min versus 4000 Hz at 60–75 min).

Overall, these patterns are aligned with the changes seen in the VENG inflammatory index profile reported in the next subsection.

Time-frequency representations of VENG, such as the ones shown here or, for example, the synchrosqueezed transform (SST),[17] can be useful in future studies when dissecting the complex patterns in VENG and relating the VENG changes to the experimental states. Future studies will explore the physiological foundations of the observed high-frequency oscillations in VENG.

### HRV and VENG inflammatory index: machine learning approach

The vagus nerve is a key homeostatic regulator of the inflammatory response. To confirm the hypothesis that the HRV inflammatory index encodes changes in VENG properties we tested whether the HRV-derived inflammatory index applied to VENG also correctly profiled the inflammatory response.

Using the same CIMVA approach that yielded the HRV inflammatory index (Fig. 5), we computed the variability of select mathematical metrics (mean rate, eScalE, sgridTau, AsymI, Multifractal_c1) for both filtered and unfiltered VENG recordings. The mathematical metrics are defined as follows: ScalE: embedding scaling exponent; sgridTau: Grid transformation feature: Time delay similarity index; AsymI: Multiscale time irreversibility asymmetry index; Multifractal_c1: MultiFractal spectrum cumulant of the first order.

We present the resulting dynamics of the computed VENG variability metrics in 5 min intervals for each experimental stage (Figure 7).

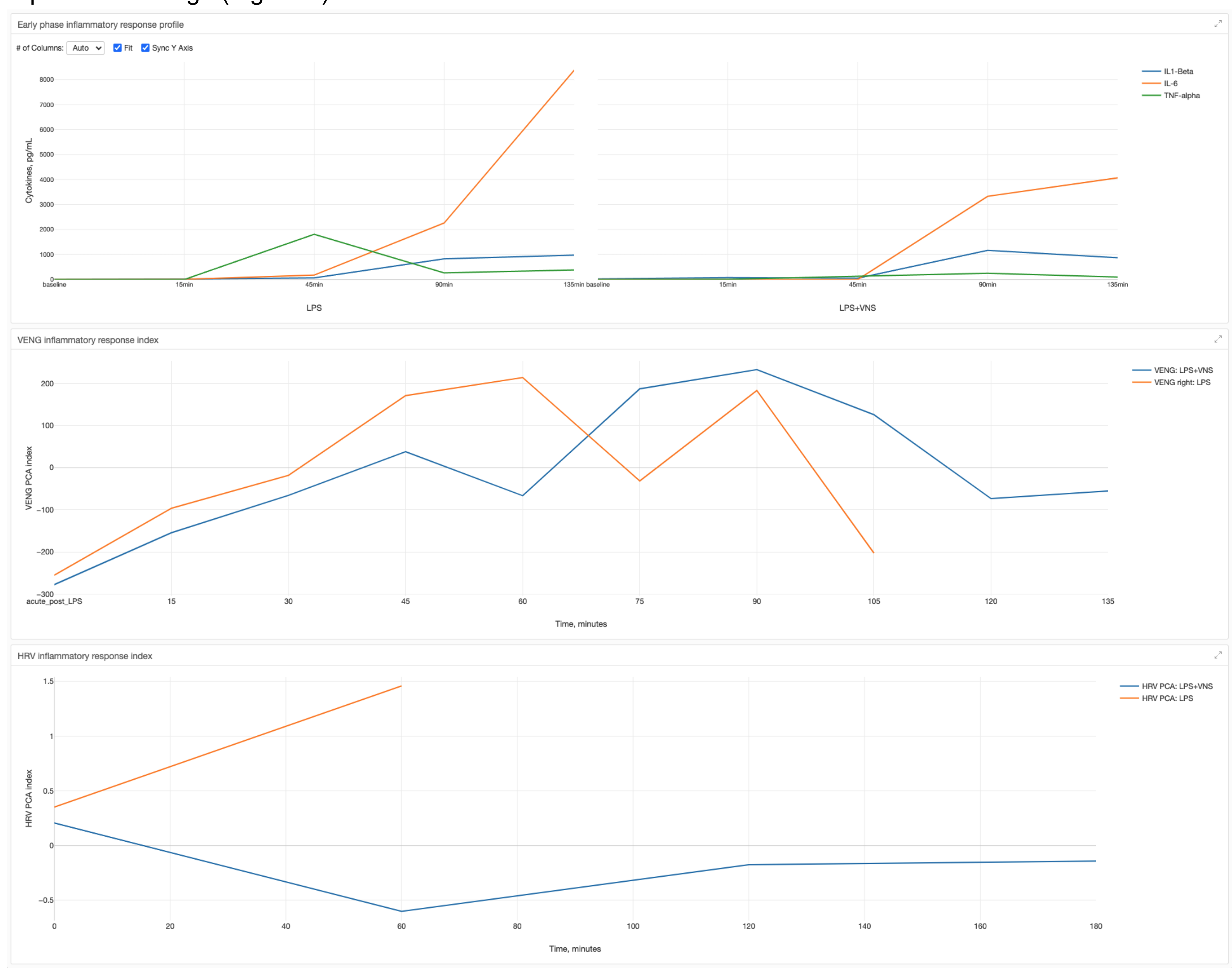

**Figure 7. Temporal profile of the fetal inflammatory index applied to VENG and HRV** together with the early stage cytokine inflammatory response**.** The interactive version is accessible here and here.

For both LPS–exposed piglets, with and without VNS treatment, there is a temporal profile of VENG-derived inflammatory index that follows that of the HRV inflammatory index and the inflammatory response almost exactly. Interestingly, in the piglet where both nerves were recorded and which received no VNS, we see side-specific differences reflecting the known functional vagus nerve asymmetry.

## Discussion

HRV inflammatory index accurately tracked the cytokines' temporal profiles, and this was reflected in the power-spectral properties of VENG in this swine model of neonatal sepsis. This approach demonstrates that the HRV inflammatory index 1) applies across two large mammalian species (sheep and swine) with strong similarities with human physiology, pre- and postnatally and 2) performs well at different degrees of sepsis (*i.e.,* nanogram and milligram doses of LPS). Moreover, the VNS paradigm based on [7] suppresses LPS-induced inflammation in neonatal piglets, even at high doses of LPS. Notably, the effects of VNS were also reflected by temporally concordant changes in the HRV inflammatory index. This suggests a certain degree of species independence in regards to the performance of the HRV inflammatory index which can be seen as a hallmark of the HRV code, a concept reviewed elsewhere in more details.[18–21]

We cannot say with certainty at this stage (having just one animal in each group) that VNS reversed leukopenia seen in the LPS piglet, but the changes are in favor of this conclusion. LPS injection triggered acid–base status changes, as well as cardiovascular and cytokine responses which were all altered by VNS treatment, albeit we did not observe an overt cardiovascular shock. The improved acid–base status is in line with a reduced inflammatory response secondary to VNS treatment. This may also explain the relatively lower, closer to physiological, blood pressure measurements under steady tachycardia in the VNS-treated piglet. The lower initial dBP levels in the VNS-treated piglet may be directly due to early VNS effects. Prolonged, but not acute, VNS has been shown to reduce dBP and mBP levels.[22–24] However, differences in VNS installation, parameter settings and species make comparison difficult requiring further systematic studies, in particular, to verify whether the early reduction of dBP is clinically relevant under conditions of neonatal sepsis.[19] In addition, the reported cardiovascular effects of VNS should be seen with restraint, since piglets can show a considerable range of blood pressure and heart rate values at this age.[11,12]

The effects of VNS on LPS-induced cytokine responses were not homogenous and showed some cytokine specificities and differences in temporal response profiles. This is in line with recent findings in rodents showing that specific VNS settings induce specific cytokine responses and, vice versa, that VENG properties encode for specific cytokine sensing in the vagus nerve.[16,25] Consequently, the selected VNS settings may have been specific to TNF-α and IL-6, but not IL-1β and IL-8: an interesting finding considering that we based the VNS settings on the work by Borovikova et al. which was done in adult rats and reduced TNF-α production.[7] In contrast to their study, we observed a transient early reduction of IL-10 in the VNS-treated piglet.

In the present study, we demonstrated that encoding the multi–dimensional properties of HRV allows for tracking the inflammatory response in real time and in parallel with the evolving cytokine release. The information we captured this way from HRV is also contained in the vagus nerve electrical activity (VENG) itself, further strengthening the notion that this experiment allows tracking of afferent/efferent brain-body neuroimmunological communication.

## Limitations

This pilot study has limitations. Only the left VENG was recorded in the VNS+LPS piglet. In the LPS piglet, as well as in future studies, bilateral VENG should be recorded to study the side differences in VENG patterns which are most likely to reflect the underlying side-specific brain-body communication. Furthermore, the control animal died in the beginning of the experiment preventing us from collecting any data on this animal. Additionally, we did not compare the effect of VNS stimulation alone (without LPS administration) on the different variables.

### Future directions

Future studies using this model could focus on recording and analyzing the VENG in the setting of septicemia. Do interventions such as VNS alter VENG properties? A similar analysis could be conducted for concomitant HRV dynamics. Next, the correlations between VENG and HRV signals and their features can be gauged. Once the number and types of different states for each signal and the correlations between them have been characterized, it could be investigated whether interventions on one signal (*e.g.*, VNS) can have the desired effect on another signal (HRV). Previous work suggests that this is possible for VENG and HRV.[26,27] Recently, promising results have been obtained in decoding VENG activity using a density-based clustering approach implemented in DBSCAN which is readily available in R.[25,28,29]

Future studies should further direct this approach to derive the mathematical properties of the VENG activity reflecting the inflammatory response which maps onto the respective mathematical properties of HRV captured by the inflammatory index. Specifically, one could ask if the properties can be predicted from VENG, which represents vagus nerve activity, i.e., a direct problem (as opposed to using HRV to predict vagus nerve activity, an inverse problem). Deep Learning approaches could be tractable. For example, one could train an LSTM network on VENG baseline data to predict the corresponding HRV features (VENG -> HRV {features set from our inflammatory index}). Future directions include also tracking image complexity changes in wavelet transform representation of the VENG, e.g., using the approach by Zhao.[30]

This research direction has the potential to expand our understanding of the HRV code through the direct study of VENG properties and direct manipulation of VENG properties by VNS. Ultimately, this will lead to closed-loop biocybernetic stimulation/monitoring systems of vagus nerve activity and its surveillance and control of the inflammatory milieu.

**Acknowledgments**

We gratefully acknowledge the highly skilled support of the Clinical Sciences team of the Veterinary Faculty at the Université de Montréal, in particular, Marco Bosa who fabricated the VNS/VENG probes. We thank Jan Hamanishi for the skillful graphical design assistance. We thank Dr. Gregory Lodygensky for clinical advice from the neonatologist's perspective during the planning stage of the study. We gratefully acknowledge the funding from CIHR and FRQS to MGF.